# Jeans instability and hydrodynamic roots of Landau damping

*Alexander Ershkovich[1] and Arnold Kiv[2]*

*1 Tel Aviv University, Ramat Aviv, Israel*

*2 Ben-Gurion University, Beer-Sheva, Israel*

Abstract

Landau damping of Langmuir waves is shown to have hydrodynamic roots, and, in principle, might have been predicted (along with Langmuir waves) several decades earlier, soon after Jeans (1902) paper appeared.

The Jeans dispersion relation is

$$\omega^2 = (kc_s)^2 - \Omega^2 = \Omega^2(\gamma k^2 D_j^2 - 1) \tag{1}$$

where $k$ is the wavenumber, $c_s^2 = \gamma v_T^2$ is the squared sound velocity, $\gamma$ is the ratio of specific heats, $\Omega = (4\pi G \rho)^{1/2}$ is the Jeans gravitational frequency, $v_T$ is the average thermal velocity, and $D_j = v_T/\Omega$ is the gravitational analog of the Debye radius (which, of course, is not associated with screening, in contrast to plasma) (Jeans 1902). Equation (1) was obtained within Hydrodynamics, by using the Jeans model for infinite, homogeneous self-gravitating gas at rest in the equilibrium. This model is under severe criticism for decades (e.g. Binney and Tremain, 2008) as allegedly inconsistent because at the equilibrium it does not obey the Poisson law for the gravitational potential. We have shown, however, (Trigger et al., 2004; Ershkovich 2011) that the Jeans model is quite self-consistent as in uniform infinite fluid at rest the gravity force vanishes due to symmetry reason, and, hence, Poisson equation in the equilibrium is just irrelevant. Well, the Jeans model, of course, is not realistic but theoretical models (sometimes extremely useful) represent an idealization which seldom is quite realistic. After all, ideal fluid, perfect gas, Maxwell demons, etc. do not exist either.

Consider now an infinite, uniform electronic gas at rest. Taking into account similarity between Newton gravitational and Coulomb laws in electrostatics, we replace the squared Jeans frequency $\Omega^2$ in equation (1) by $-\omega_0^2$, where $\omega_0$ is the Langmuir (1926) frequency, and obtain the dispersion relation for Langmuir waves

$$\omega^2 = \omega_0^2 \left(3k^2 D_e^2 + 1\right) \qquad (2)$$

where $D_e = v_T / \omega_0$ is the Debye radius of electronic gas. Equation (2) also may be derived for plasma in hydrodynamic approximation (e.g. Krall and Trivelpiece, 1973), one-dimensional treatment when $\gamma = 3$ is justified for Langmuir waves (Krall and Trivelpiece, 1973). Equation (2) yields a real part of the wave frequency $\omega$ in the kinetic treatment of the Langmuir waves (Landau, 1946; see also Krall and Trivelpiece, 1973). The Jeans model above for dusty plasma has been considered in Ershkovich and Israelevich (2008) by using Boltzmann-Vlasov equation in kinetics.

Consider now a simplified picture: three identical particles (charged or not) are located along the same line at the points *A*, *B*, and *C* (so that in the equilibrium the distances *AB* = *BC*). A small sporadic displacement of the particle *B* toward the particle *C* violates the balance of forces (gravitational or electrostatic), because the distance between particles *B* and *C* becomes less than between *A* and *B*. As a result, an interaction between particles *B* and *C* grows, and between *A* and *B* diminishes. In case of self-gravitation, due to violation of symmetry, a particle *B* continues to approach the particle *C* (it corresponds to Jeans instability) whereas Coulomb repulsion tends to return a charged particle B at its original position, with small oscillation around the equilibrium state (with the frequency $\omega_0$). Chaotic thermal motion, naturally, tends to destroy this idealized picture, hinders an organized, regular motion (waves and instability), giving rise to collisionless damping of oscillations. This effect is illustrated by equation (1): when thermal velocity is small ($\gamma k^2 D_j^2 \ll 1$) the increment of the Jeans instability ($\text{Im}\,\omega = \Omega$) is maximal. If the thermal velocity $v_T$ grows, the increment diminishes, and with $\gamma v_T^2 = \Omega^2 / k^2$ ($kD_j \sim 1$) the instability vanishes for any finite value of the wave number *k*. An organized motion is suppressed by the chaotic one. The damping is caused by thermal motion. Indeed, equation (2) points to the same conclusion: with $(kD_e)^2 \gg 1$ equation (2) describes thermal oscillations of the electronic gas:

$\omega/k \approx \sqrt{3} v_T$ (i.e. electronic acoustic waves), whereas Langmuir waves almost disappear (this case corresponds to strong collisionless damping of these waves). In the opposite limit, $(kD_e)^2 \ll 1$, equation (2) reduces to $\omega \approx \omega_0$. This is a condition of Landau damping. Thus, results of hydrodynamic and kinetic treatments happen to be qualitatively very similar, namely, strong damping of Langmuir waves within the Debye sphere and small damping outside of it. Moreover, the central idea of resonance between particles and fields, with the energy exchange between them, resulting in collisionless energy dissipation is implicitly present in equation (2), being associated with relation $(kD_e)^2 \sim 1$, that is $\omega/k \sim v_T$ (which is indicative of resonance). From here it is not far to Landau (1946) damping (with its interpretation as reverse Cherenkov effect). Of course, some questions cannot be answered by means of hydrodynamic approximation, for instance, what happens with $(kD_e)^2 \approx 1$? If chaotic thermal motion causes the wave damping what is the decrement? These questions may be elucidated only by means of exact kinetic approach. Landau damping of Langmuir waves proportional to $\exp(\text{Im}\,\omega t)$, $\text{Im}\,\omega < 0$ was obtained by Landau (1946) with $(kD_e)^2 \ll 1$, $|\text{Im}\,\omega| \ll \text{Re}\,\omega$ (see also, e.g., Krall and Trivelpiece, 1973). Of course, Landau damping is of tremendous heuristic value. This effect is generally (and fairly) believed to be a corner-stone, an axiom of plasma kinetics. We have to say, however, that the Landau (1946) rule of a pole bypass direction was chosen in Landau (1946) *ad hoc*, as if specially in order to describe mathematically a damping (rather than instability). An assumption of Maxwell distribution seems also not to be quite well consistent with the collisionless character of damping.

The effect of Landau damping of Langmuir waves is very small ($|\text{Im}\,\omega| \ll \text{Re}\,\omega$). According to Ecker (1972), it may be experimentally observed only in a very narrow interval $0.2 \leq kD_e \leq 0.4$.

To resume, the chaotic thermal motion must hamper the regular, organized motion like waves and instabilities resulting in wave damping. This effect is seen from equation (1) for Jeans instability and should also be present in equation (2) for Langmuir waves. Therefore we believe that our qualitative picture above (including Langmuir waves and their collisionless damping), in principle, might have been predicted several decades earlier, soon after Jeans (1902) pioneering paper appeared.

**Conclusion**

Thermal motion was shown to suppress the regular one, thereby resulting in wave damping. Kinetic treatment also supports this (almost obvious) statement. But, according to Landau (1946) (see also Krall and Trivelpiece (1973)) both $\text{Re}\,\omega$ and $\text{Im}\,\omega$ depend on $kv_T$ terms. We remind that $\text{Re}\,\omega$ obeys the dispersion relation (2) which is obtained in hydrodynamical approximation. The origin of thermal terms (depending on $kv_T$) both in $\text{Re}\,\omega$ and $\text{Im}\,\omega$ is the same: it is the distribution function. Thus, there is genetic connection between Landau damping and Hydrodynamics. The arguments above allow us to arrive at the conclusion that this damping might have been predicted (of course qualitatively) long before the plasma kinetics arose. Landau damping is generally believed to be purely kinetic effect. To our opinion, it looks similar to the statement that bacteria arose together with Leeuwenhoek microscope.

Trigger, S., A. Ershkovich, G. van Heijst, and P. Schram, *Phys. Rev.* **E69**, 066403, 2004